\documentclass[prl, aps, twocolumn, groupedaddress,showpacs]{revtex4}
\usepackage{amsmath}
\usepackage{graphicx}
 
\begin{document}

\title{Chirality of topological gap solitons in bosonic dimer chains}
\author{D. D. Solnyshkov, O. Bleu, B. Teklu, G. Malpuech}
\affiliation{Institut Pascal, PHOTON-N2, University Clermont Auvergne, CNRS, 4 avenue Blaise Pascal, 63178 Aubi\`{e}re Cedex, France.} 

\begin{abstract}
We study gap solitons which appear in the topological gap of 1D bosonic dimer chains within the mean-field approximation. We find that such solitons have a non-trivial texture of the sublattice pseudospin. We reveal their chiral nature by demonstrating the anisotropy of their behavior in presence of a localized energy potential.
\end{abstract}

\maketitle

Topologically non-trivial structures are currently in the focus of attention of scientific community. Topological insulators are studied in electronic systems for fermionic particles \cite{Hasan2010}, but also in analog systems for bosonic particles (atomic lattices and photonic "topological mirrors" \cite{Satija2013,aidelsburger2015measuring,Schomerus2013,Lu2014,Poddubny,Poddubny2015,Slobozhanyuk2015,Sinev2015}). The advantage of artificial photonic systems lies in their design flexibility and the possibility of direct wavefunction measurements. The properties of such structures are relatively well explored in the linear regime, where the topological invariants have been found to characterize the bands \cite{Simon1983} and determine their properties, including the existence of chiral edge states \cite{Hatsugai1993}. The nonlinear regime is much less explored.
Indeed, an interacting quantum fluid exhibits topological properties on its own \cite{VolovikReview}, and one can expect them to become even richer when combined with the topology of the dispersion in the linear case \cite{Lumer2013,furukawa2015excitation,engelhardt2015topological,Bleu2016,
xu2016pi,di2016topological}.

A 1-dimensional (1D) periodic lattice with a certain degree of dimerization is one of the simplest lattices exhibiting topological properties \cite{Ryu2010,Asboth2016,Guo2016}. Such structure shows a splitting of a single $s$-type band into two bands, corresponding to the bonding and anti-bonding states of the individual dimers. These subbands are separated by a gap, characterized by a topological invariant -- the Zak phase \cite{Zak1989}. The properties of nonlinear solutions existing in this gap -- the gap solitons -- can be expected to be strongly modified with respect to the solitons in the ordinary gap. The Su-Schrieffer-Heeger soliton is perhaps one of the most famous examples of topologically nontrivial solutions \cite{Su1980} for a dimer chain. However, it involves dynamical dimerization, that is, modification of the properties of the lattice itself: this soliton is a domain wall between two distinct lattices. Similar dimerization domains can be observed in ionic chains \cite{delCampo2010,Pyka2013} and artificially created in photonic chains \cite{Segev2016}. Recently, chiral solitons of the SSH type were observed in double chains \cite{Cheon2015}.
But there also exist solitonic non-linear solutions that do not require the modification of the lattice. Many of them have been studied in dimerized and zigzag lattices in acoustics \cite{Kivshar1992}, Bose condensates \cite{Haddad2015}, and photonic systems \cite{Sukhorukov2002,Efremidis2002,Vicencio2009,Dmitriev2010} (including PT-invariant ones \cite{Musslimani2008,Suchkov2011,Belicev2012,Wimmer2015,Suchkov2016}), with a particularly interesting recent experimental observation \cite{Kanshu2012}.  However, the crucial role played by the anisotropy of the Bloch part of the soliton wave function with respect to the two different atoms forming the lattice (and defining the sublattice pseudospin) has remained unnoticed.

In this work, we demonstrate that a gap soliton in a single dimer chain can exhibit chirality. We study a gap soliton in the topological gap of a dimer chain, first using the variational approach, and then by direct solution of the Gross-Pitaevskii equation with a periodic potential. This solution is strongly different from the SSH soliton \cite{Cheon2015}, because it does not involve the modification of the lattice itself. It is also different from the dark-bright solitons \cite{Solnyshkov2016}, because it does not involve neither the polarization degree of freedom, nor an extended condensate. The topological gap soliton (TGS) is a typical localized solution, appearing from negative mass states at the boundary of a topological gap. We demonstrate that such solitons exhibit a nontrivial pattern of sublattice pseudospin due to their negative mass and pseudospin-anisotropic interactions. We determine their sublattice-polarization degree and demonstrate the chiral nature of these solitons via their asymmetric behavior, which gives a striking contrast with the isotropic behavior of non-topological gap solitons (GS). These results are confirmed by direct calculations. 

The practical realization of the system can be based on a patterned microcavity in the regime of strong coupling \cite{Tanese}, with the single-particle states being the cavity exciton-polaritons, hybrid light-matter particles characterized by strong interactions thanks to their excitonic fraction, and where bright solitons are observed even without patterning \cite{Egorov2009,Sich2012}. However, our results are valid for any photonic system, where solitonic states can be observed thanks to non-linearities, and also for atomic condensates, for which periodic lattices are routinely created \cite{Pitaevskii}, but which would require putting the condensate out of thermal equilibrium. A closer look at recent experimental data in a photonic dimer chain \cite{Kanshu2012} confirms our predictions for the chiral nature of the TGS.

\begin{figure}[tbp]
\includegraphics[scale=0.6]{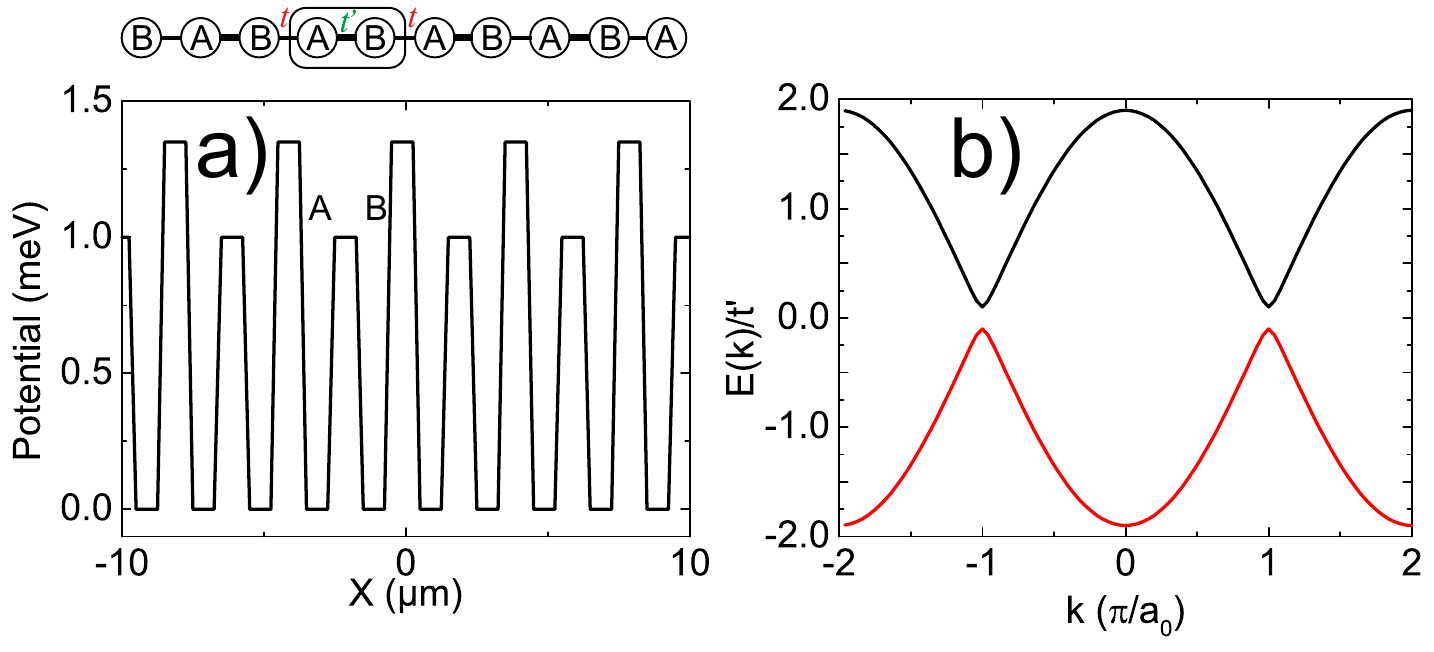}
\caption{ (Color online) a) Periodic potential of a dimer chain and the corresponding tight-binding representation. b) Tight-binding dispersion of the dimer chain with the topological gap in the middle for $t=0.9t'$. }
\label{figpot}
\end{figure}

\emph{The model}. We begin with the tight-binding description of the dimer chain shown in Fig. \ref{figpot}(a). Each minimum of the potential corresponds to an individual site, which is called A or B. Since the barriers between the sites have different heights, the tunneling coefficients $t$ and $t'$ are also different.
If one neglects the degree of freedom corresponding to the polarization of light or the spin of electrons, the Hamiltonian of a dimer chain in the tight-binding approximation can be written as \cite{Delplace2011}:

\begin{equation}
\hat H = \sum\limits_{m = 1} {t'\hat b_m^\dag {\hat a_m} + t\hat a_{m + 1}^\dag {\hat b_m}}  + H.c.
\end{equation}
where $\hat a,\hat b$ are the annihilation operators on the corresponding atoms (A and B, Fig. \ref{figpot}(a)).  We assume that $t'>t$, meaning that the unit cell A-B corresponds to a tightly bound "molecule".
Using the Bloch theorem, this Hamiltonian can be rewritten in the matrix form in the basis $\psi_k=\left(\psi_{A,k},\psi_{B,k}\right)^T$:

\begin{equation}
\hat H\left( k \right) =  - \left( {\begin{array}{*{20}{c}}
0&{t' + t{e^{ - ika_0}}}\\
{t' + t{e^{ika_0}}}&0
\end{array}} \right)
\label{fullham}
\end{equation}
with period $a_0$. The sublattice spinor $\left(\psi_{A,k},\psi_{B,k}\right)^T$ allows defining the sublattice pseudospin: $S_Z=(|\psi_A|^2-|\psi_B|^2)/2$, $S_X=\Re (\psi_A\psi_B^*)$, $S_Y=\Im (\psi_A^*\psi_B)$. The Hamiltonian can then be represented as an effective magnetic field $\mathbf{\Omega}(k)$ acting on this pseudospin $H=-\hbar\mathbf{\Omega}\mathbf{S}/2$.

 The dispersion of the chain is given by
\begin{equation}
{E_ \pm }\left( k \right) =  \pm \sqrt {{t^2} + t{'^2} + 2tt'\cos \left( {ka_0} \right)} 
\end{equation}
It is plotted in Fig. \ref{figpot}(b). 
The topological invariant analog, characterizing the two subbands, is the Zak phase \cite{Zak1989}. Contrary to the Chern number, the Zak phase is gauge-dependent \cite{Atala2013}: the unit cell of a chain with inversion symmetry can be chosen both for $t'>t$ and $t'<t$ in such a way \cite{Zak1989,Kohn1959} that the Zak phase of a given band is $\pm\pi$, indicating nontrivial topology (associated with protected edge states in finite chains) induced by the dimerization \cite{suppl}. The gap between these bands, appearing due to dimerization, can thus be called "topological".

The TGS is a stable localized solution of the nonlinear equation, whose energy lies in the topological gap. An ordinary GS with its energy above the upper allowed band can also appear in the same lattice. We are going to study the properties of the TGS and compare it with the ordinary GS.
To find the non-linear soliton solution, we use the variational approach. The gap solitons are usually formed from the Bloch states  at the edge of the gap. These Bloch states will determine the wavefunction of the soliton: for TGS, the wavefunction changes sign between the unit cells ($k=\pi/a_0$), whereas for the ordinary GS (upper gap), the wavefunction changes sign between each pillar ($k=2\pi/a_0$). This is why the TGS was also called antisymmetric soliton \cite{Efremidis2002}. The most important feature of the gap soliton, made of negative mass states, is that it has to maximize the energy, and not to minimize it.

In our dimer chain, the trial function has to take into account the fact that the interactions are spin-anisotropic with respect to the sublattice pseudospin. Indeed, a particle on a given site (say, A) interacts only weakly with a particle on a different site (say, B). Maximal interaction energy sought by the soliton is therefore achieved by putting all particles on the same lattice site, that is, by the "circular" polarized states of the sublattice pseudospin, and the corresponding "effective field" is oriented in the $Z$ direction. The pseudospin cannot be constant everywhere, because other terms in the Hamiltonian (appearing due to dimerization) correspond to fields in the $X$ and $Y$ directions (see \cite{suppl} for details). We can thus expect the soliton pseudospin texture to be non-trivial, as a consequence of the gap topology.

A general shape of the trial function with the two pseudospin components can be constructed using the hyperbolic secant profile, known to be a good solution for the bright soliton of the Gross-Pitaevskii equation:
\begin{equation}
{\psi}\left( {x,a,b} \right) = 2\sqrt {n/a} \left( {\begin{array}{*{20}{c}}
{1/\cosh \left( {\left( {x - b} \right)/a} \right)}\\
{1/\cosh \left( {\left( {x + b} \right)/a} \right)}
\end{array}} \right)
\end{equation}
where $a$ is the soliton width, $b$ is the displacement of the maximum of each component with respect to the global center of mass, and $n$ is the soliton density. Close to the edge of the Brillouin zone, the Hamiltonian is reduced to the Dirac equation with nonlinear terms, extensively studied in the past \cite{Werle1977,Takahashi1979,Bartsch2006,Pelinovsky2012,Haddad2015}. However, it does not have stable solutions in our case, because of the pseudospin-anisotropic interactions (see \cite{suppl} for details). Thus, we consider the full Hamiltonian \eqref{fullham} in the reciprocal space and work with the Fourier transforms of the trial wavefunctions to calculate the kinetic energy $E_{kin}(a,b)$. 
To calculate the interaction energy, the integration should be performed in real space:
\begin{equation}
E_{int}(a)=\frac{1}{2}\alpha \int\limits_{ - \infty }^{ + \infty } {\left( {{{\left| {{\psi _A}} \right|}^4} + {{\left| {{\psi _B}} \right|}^4}} \right)dx}
\end{equation}
which gives a $1/a$ dependence $E_{int}=n^2/12a$. 
\begin{figure}[tbp]
\includegraphics[scale=0.375]{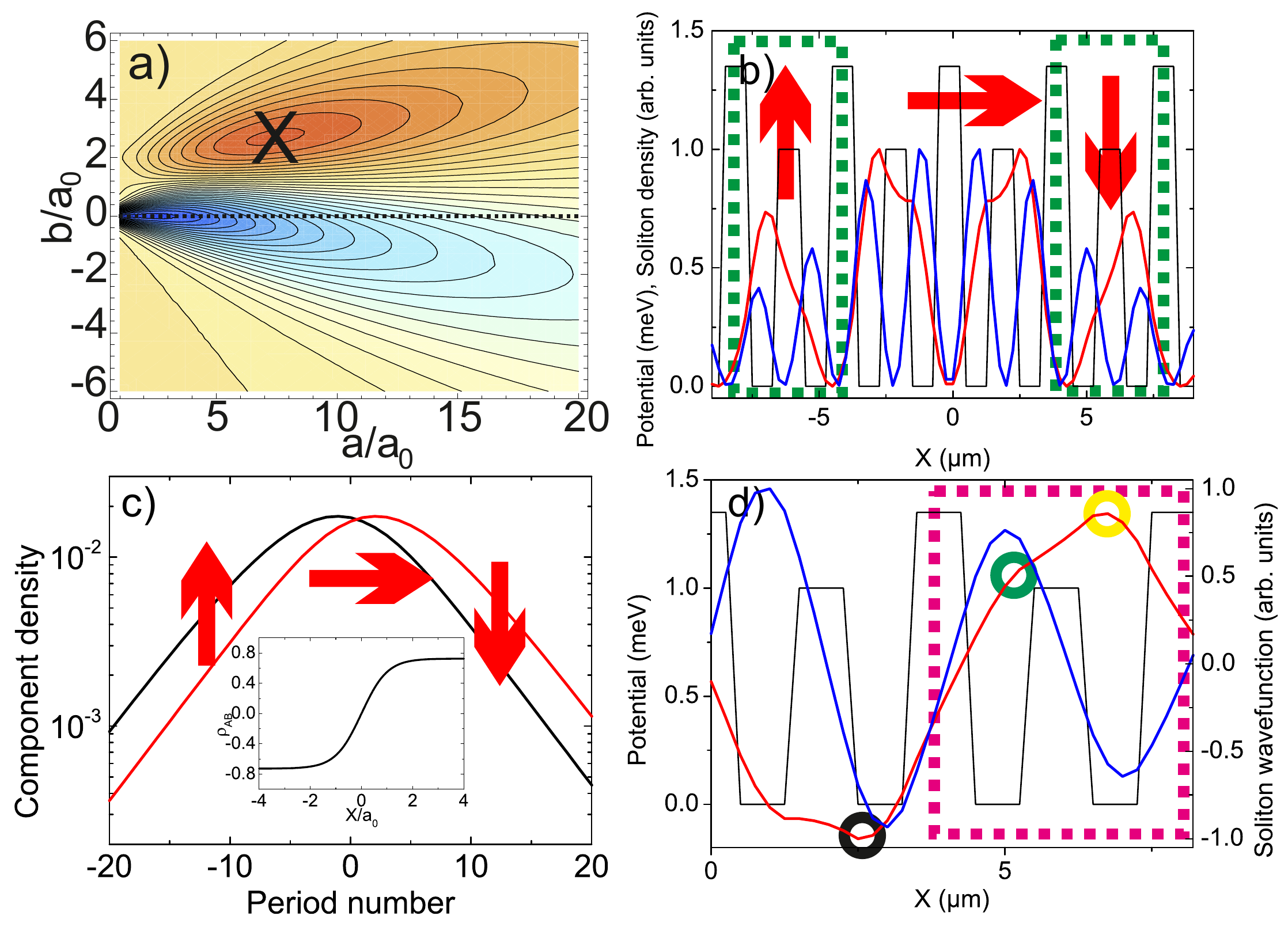}
\caption{ (Color online) a) The energy as a function of variational parameters $a,b$ (red - maximum); b) Potential profile together with the GS (blue) and TGS (red) density, demonstrating opposite TGS sublattice polarization (red arrows) on two cells (marked in green); c) The two sublattice pseudospin components (A - black, B - red) extracted from the full wavefunction $\psi$. Inset shows the sublattice polarization degree; d) Potential and the GS (blue) and TGS (red) wavefunctions (circles discussed in the text). }
\label{figenpolar}
\end{figure}

The variational energy $E_{var}(a,b)=E_{kin}(a,b)+E_{int}(a,b)$ demonstrates a local maximum with respect to both $a$ and $b$, as shown in Fig. \ref{figenpolar}(a). The anisotropy of the TGS is clearly visible in this figure: a maximum (marked with a cross) is present only for a positive value of $b$ (determined by the dimerization of the lattice), corresponding to a particular pseudospin texture, whereas the other pseudospin texture does not allow a stable solution (see \cite{suppl} for details). Therefore, the TGS indeed has a nontrivial pseudospin texture.

To verify the analytical solution, we have solved the Gross-Pitaevskii equation for a realistic periodic 1D square potential, which can be obtained for photons by patterning a wire cavity \cite{Tanese} or by working with coupled waveguides \cite{Kanshu2012,Segev2016}, or for bosonic atoms in an optical lattice \cite{Satija2013}. The solution on a grid (without the tight-binding approximation) is obtained by using the iterative method. The equation reads:
\begin{equation}
E\psi \left( x \right) =  - \frac{{{\hbar ^2}}}{{2m}}\frac{{{\partial ^2}}}{{\partial {x^2}}}\psi \left( x \right) + \alpha {\left| {\psi \left( x \right)} \right|^2}\psi \left( x \right) + U\left( x \right)\psi \left( x \right)
\label{GPE}
\end{equation}
 
Here $U(x)$ is the periodic potential of a dimer chain, shown in Fig. \ref{figpot}(a). This equation does not contain the sublattice pseudospin in the explicit way ($\psi$ is not a spinor), because it is not in the tight-binding approximation. However, the pseudospin can be extracted from the solution $\psi(x)$ by analyzing the densities in even and odd minima of the potential separately: $n_{A/B}(x)=\int |\psi(x)|^2U_{A/B}(x)\,dx$. Next, we study the internal structure of GS and TGS more in detail to verify our predictions.

Figure \ref{figenpolar}(b) highlights the two opposite sides of the TGS (red curve), where its sublattice polarization (red arrows) is clearly visible. We see that, counter-intuitively, at the left edge the intensity is mostly concentrated on the A atoms ("spin"-up), whereas on the right edge the intensity is on the B atoms ("spin"-down), contrary to the GS (blue), showing a typical soliton profile. This feature is present in calculated and measured figures of Refs. \cite{Efremidis2002,Vicencio2009,Kanshu2012}, but it has not drawn the attention it deserves as a signature of anisotropy of the soliton. The extracted density of each sublattice pseudospin component (black, red) is shown in Fig. \ref{figenpolar}(c). The log scale  plot clearly exhibits a $1/\cosh^2(x/a)$ dependence of a bright soliton, with the two components displaced with respect to the soliton center, justifying our trial wavefunction.

The variational approach  allows us to find the sublattice-polarization degree $\rho_{AB}$ of the TGS, which is the density difference between the $A$ and $B$ sites (see the inset of  Fig. \ref{figenpolar}(c)), 
\begin{equation}
\rho_{AB}\left(x\right)=\tanh{\left(b/a\right)}\tanh{\left(x/a\right)}
\end{equation}
which, considering the limit $x\to \infty$, gives $\rho_{AB\infty}=\tanh{b/a}$.
This result characterizes the sublattice-polarization texture of the gap soliton, and since the soliton size $a$ decreases with the number of particles while $b$ remains fixed, its polarization degree has to increase with $n$.

The counter-intuitive TGS density distribution within the unit cell, with more particles on the pillar away from the soliton center (contrary to the usual GS), can be understood qualitatively from Fig. \ref{figenpolar}(d), showing the wave function over two unit cells. To maximize the interaction energy, the particle distribution within each dimer (rose rectangle) should be maximally anisotropic, and the particles tend to localize either on A (green circle) or B (yellow circle). On the other hand, the main contribution to the kinetic energy is due to the change of sign of the wave function between the cells. Its minimization imposes the wave function to be minimal on the A pillar (green circle), because the neighboring cell is closer to TGS center and thus has higher density (black circle) than the other neighbor. This is in contrast with the ordinary GS, which has a wavefunction changing sign between \emph{each} pillar, and is therefore not subject to this polarization mechanism.

\begin{figure}[tbp]
\includegraphics[scale=0.35]{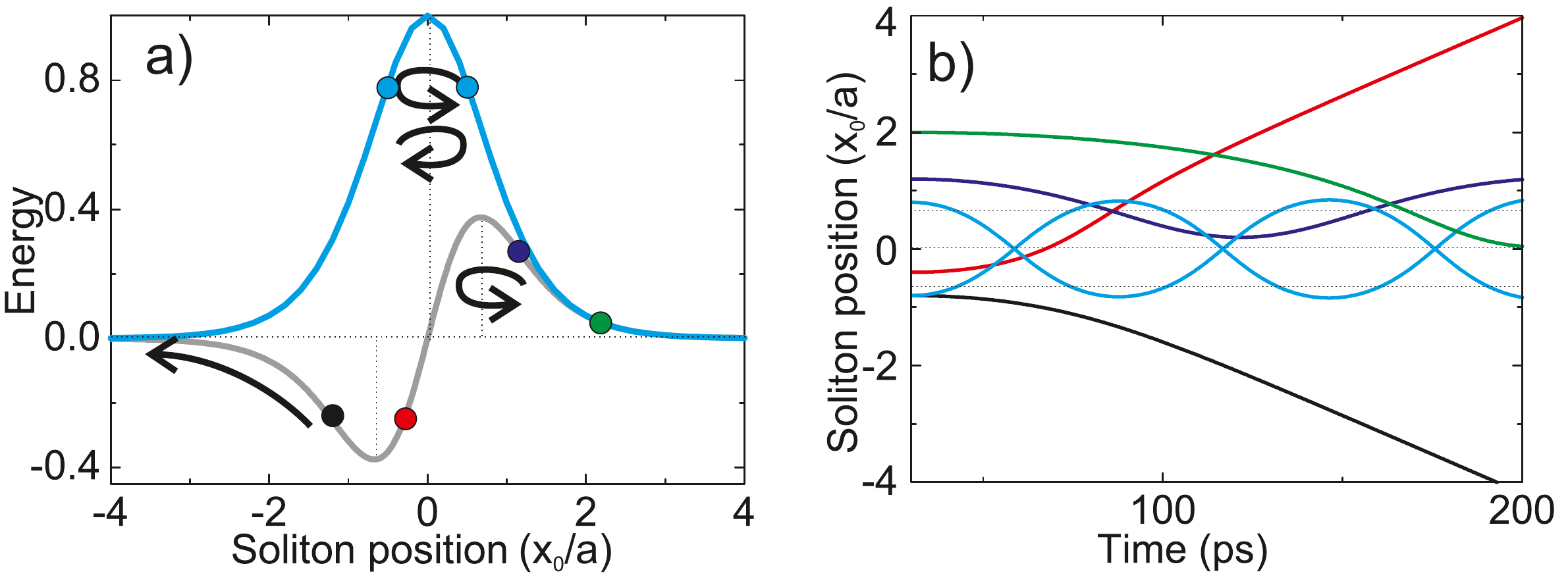}
\caption{ (Color online) a) Energy of the soliton as a function of its position with respect to the localized potential $\delta(x)$: TGS (gray) and GS (blue). b) Soliton trajectories of different families: TGS (black, red, navy, green) and GS (cyan).}
\label{figpseudo}
\end{figure}

The opposite polarization degree of the sublattice pseudospin on each side is crucial, because it distinguishes the TGS from the GS of the upper gap and leads to the anisotropic behavior of TGS. It can be experimentally probed by considering the effect of a localized potential breaking the symmetry between the A and B sites. In the tight-binding approximation, such potential can be expressed as a local effective magnetic field $\Omega_Z=\delta\left(x\right)$, and the energy of the soliton centered at $x_0$ in the presence of such field is given by
\begin{equation}
{E_{TGS}} = \int {\mathbf{\Omega}  \cdot \mathbf{S}\,dx} \propto \frac{{\tanh\left(b/a\right)\tanh \left( {{x_0}/a} \right)}}{{{{\cosh }^2}\left( {{x_0}/a} \right)}}
\end{equation}
The asymmetry of this expression is seen in Fig. \ref{figpseudo}(a) (gray line). A TGS located on one side of the field will be attracted to the defect, whereas a TGS located on the other side will be repelled to infinity as indicated by the black arrows. On the contrary, the energy of the ordinary GS formed from the states of the upper band in the presence of a $\delta$-potential can be written as:
\begin{equation}
{E_{GS}} = \int {V\left(x\right) \left|\psi\left(x\right)\right|^2\,dx} \propto \frac{1}{{{{\cosh }^2}\left( {{x_0}/a} \right)}}
\end{equation}
It is plotted in Fig. \ref{figpseudo}(a) (blue line): a positive localized potential attracts the ordinary GS whatever its initial position, which oscillates around this defect.

We have calculated the dynamics of both TGS and GS solving Hamilton's equations:
\begin{equation}
\dot x_0 = \frac{{\partial H}}{{\partial p_0}},\,\dot p_0 =  - \frac{{\partial H}}{{\partial x_0}}
\end{equation}
where $x_0$ and $p_0$ are the TGS/GS position and momentum, respectively, and its Hamiltonian is 
\begin{equation}
H(x_0,p_0)=\frac{p_0^2}{2m}+E_{TGS/GS}\left(x_0\right)
\end{equation}
where $m$ is the soliton mass. We take $p_0(t=0)=0$ as an initial condition. The resulting soliton trajectories can be classified into several families, depending on the initial position $x_0(t=0)$ and on the soliton type, shown in Fig. 3(b). The ordinary GS trajectories are shown in cyan, for the initial positions shown as cyan points on Fig. \ref{figpseudo}(a). The GS is always confined and exhibits anharmonic oscillations because of the potential profile $E_{GS}\propto 1/\cosh^2(x_0/a)$. The TGS can be either confined (blue and green lines, initial positions in blue and green on the Fig. 3(a)), or delocalized (black and red). The regime depends on the sign of the TGS energy determined by its initial position $x_0(t=0)$. The period of the anharmonic oscillations for the localized case strongly depends on the energy (compare blue and green curves).

\begin{figure}[tbp]
\includegraphics[scale=0.27]{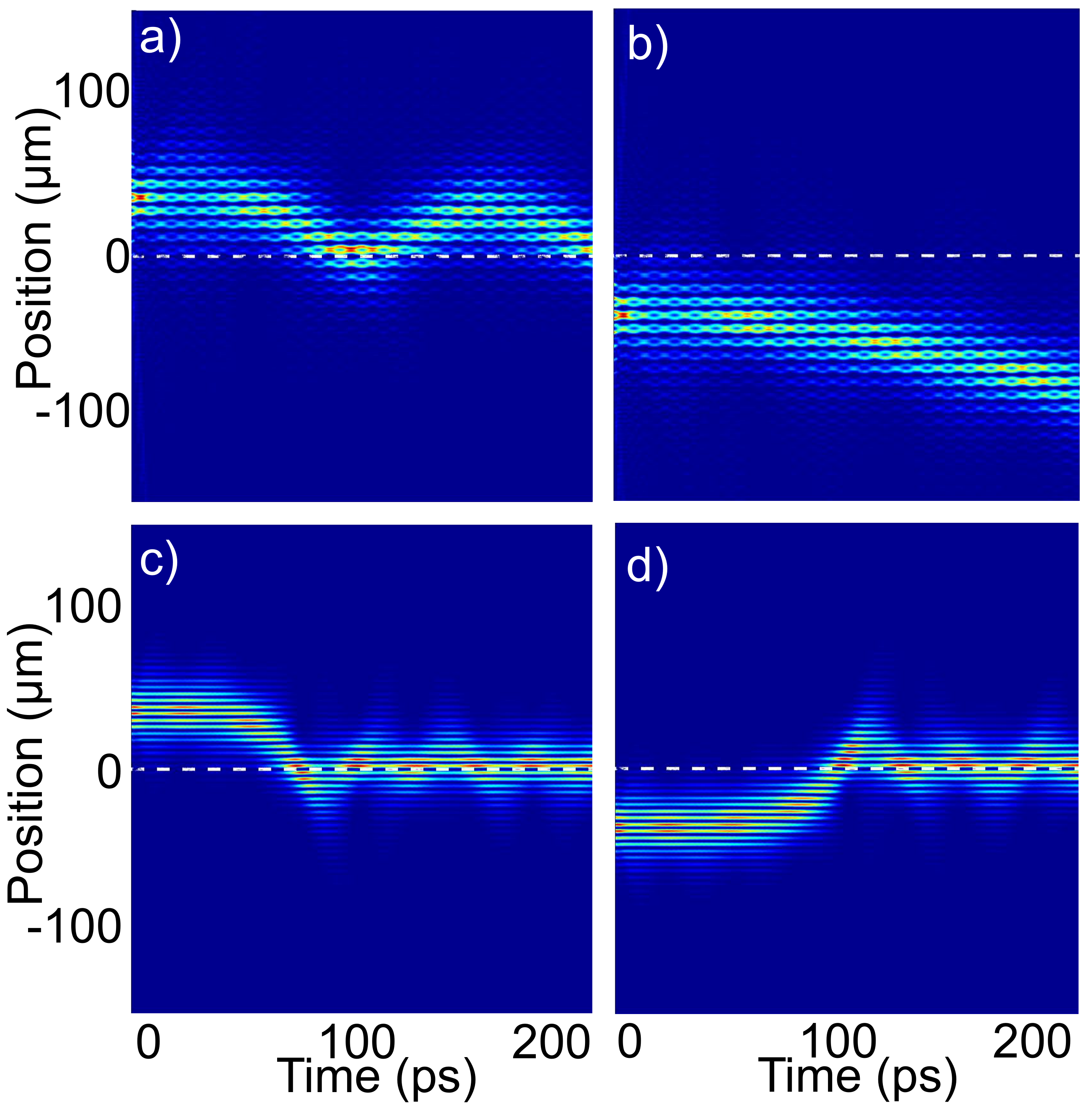}
\caption{ (Color online) Soliton trajectories plotted as the particle density as a function of position and time: a,b) TGS, oscillating trajectory or free acceleration, depending on the initial soliton position. c,d) oscillating trajectory of an ordinary GS for the same defect.
}
\label{solitons}
\end{figure}

This behavior, which is the main dynamical consequence of the TGS chirality, is confirmed by numerical simulations, shown in Fig. \ref{solitons}, performed by solving the time-dependent Gross-Pitaevskii equation 
\begin{equation}
i\hbar \frac{{\partial \psi }}{{\partial t}} =  - \frac{{{\hbar ^2}}}{{2m}}\Delta \psi  + \alpha {\left| \psi  \right|^2}\psi  + U\psi
\label{GPET}
\end{equation}
for polaritons (see \cite{suppl}) with a pulsed excitation 
\begin{equation}
{\left. {\psi \left( x \right)} \right|_{t = 0}} = \sqrt n {e^{ - {{\left( {x - {x_0}} \right)}^2}/{\sigma ^2}}}\sin \left( {\frac{{2\pi x}}{{{a_0}}}} \right)\cos \left( {\frac{{\pi x}}{{{a_0}}}} \right)
\end{equation}
for TGS and without the cosine for GS. 
Depending on the initial position, the TGS is either attracted to the point-like magnetic field, in which case it oscillates (Fig. \ref{solitons}(a)) or repelled and accelerated away from it (Fig. \ref{solitons}(b)). This behavior is a clear signature of its anisotropy, manifested in the pseudospin pattern. Contrary to the TGS, an ordinary GS does not exhibit this anisotropic behavior (oscillating behavior in both Fig. \ref{solitons}(c) and (d)), because it does not have the chiral pseudospin texture. The  agreement between the analytical model based on the Hamilton's equations and the full numerical simulations can be seen in Suppl. material (\cite{suppl}, Fig. S2).

To conclude, we have analyzed the properties of solitons in the topological gap of a 1D bosonic dimer chain. We have found that such solitons exhibit a chiral pattern of their sublattice pseudospin, allowing them to behave anisotropically, contrary to ordinary GS. Analytical solution for the soliton shape and pseudospin is confirmed by numerical simulations, including those of previous works \cite{Efremidis2002,Vicencio2009} and by a recent experimental observation (Fig. 2(b) of Ref. \cite{Kanshu2012}), but the pseudospin texture and the chiral behavior of the antisymmetric soliton have passed unnoticed in these works. 

We acknowledge the support of the ANR "Quantum Fluids of Light".

\section{Supplemental Material}

\subsection{On the topology of a Dimer chain}
The Zak phase is not as good a topological invariant as the Chern number: it is gauge-dependent and its value is strongly linked with the choice of the unit cell \citep{Atala2013}, because the integration in 1D cannot be carried out over a closed loop. However, a difference in the Zak phase between two topological phases does not depend on this choice, and indicates the non-trivial topology for both of them.

For an infinite chain, there is no difference between the cases with $t>t'$ and $t<t'$, because the unit cell can always be chosen to obtain any value of the Zak phase \cite{Zak1989}, including $\pm\pi$. However, once the unit cell is fixed, the difference between the Zak phases for the situations with $t>t'$ and $t<t'$ will always be $\pi$.

For a finite chain, once the unit cell is fixed, everything depends on the termination of the chain: if the terminating atoms are weakly coupled (correspond to $t$ or $t'$, whichever is smaller), there will be the associated topological edge states. That is, even for the choice of the unit cell of \cite{Delplace2011} which gives Zak phase 0 for the "trivial" case of $t'>t$, for which the authors do not find edge states, it is still possible to have the edge states if one removes 1 more atom from each edge of the chain. Therefore, both phases are topological, because both can have edge states \cite{Atala2013}. For a simple (non-dimerized) chain, such states are absent, and the chain is therefore trivial.

\subsection{Dirac Hamiltonian}
Since the gap soliton is formed from the states close to the edge of the Brillouin zone, it is logical to consider the Hamiltonian only at small wavevectors $q$, so that $k=\pi/a_0+q$, which gives rise to the massive Dirac equation, if one keeps only the 1st order terms:

\begin{equation}
\hat H_D =  - \left( {t' - t} \right){\sigma _X} + tqa_0{\sigma _Y}
\end{equation}
where $\sigma_X$ and $\sigma_Y$ are the Pauli matrices. We see that the difference of the tunneling coefficients $t$ and $t'$ induces a constant in-plane effective magnetic field $t'-t$ in the X direction, equivalent to a mass. Without the nonlinear term, one could redefine the pseudospin axes to obtain the more traditional shape of the Dirac equation with the mass term entering with $\sigma_Z$. However, in our case it is important to keep the original definition of the pseudospin to be able to add the interactions into the equation.

Indeed, the interactions should appear as a nonlinear term. The nonlinear Dirac equation has already been considered in previous works\cite{Takahashi1979,Bartsch2006,Pelinovsky2012}. However, the interactions in these previous works were different. Indeed, if contact interactions are taken into account in the particular case of a dimer chain, particles on different sites do not interact with each other, and the interactions become spin-anisotropic in the lattice pseudospin basis, entering the equation with $\sigma_Z$. The corresponding equation for the evolution of the wavefunction reads:

\begin{equation}
i\hbar \frac{\partial }{{\partial t}}\left( {\begin{array}{*{20}{c}}
{{\psi _A}}\\
{{\psi _B}}
\end{array}} \right) = {\hat H_D} \left( {\begin{array}{*{20}{c}}
{{\psi _A}}\\
{{\psi _B}}
\end{array}} \right) + \alpha \left( {\begin{array}{*{20}{c}}
{{{\left| {{\psi _A}} \right|}^2}{\psi _A}}\\
{{{\left| {{\psi _B}} \right|}^2}{\psi _B}}
\end{array}} \right)
\end{equation}
Let us first analyze the situation without taking account the lattice pseudospin, by choosing identical trial functions for A and B atoms.
Applying variational approach to the typical Gaussian bright soliton test function
\begin{equation}
\psi \left( {x,a} \right) = \frac{\sqrt{n}}{{{{\left( {2a} \right)}^{1/2}}{\pi ^{1/4}}}}\exp \left( { - \frac{{{x^2}}}{{2{a^2}}}} \right)\left( {\begin{array}{*{20}{c}}
1\\
1
\end{array}} \right)
\end{equation}
symmetric with respect to the components, with the variational parameter $a$ gives the following results for the kinetic energy and interaction energy contributions:
\begin{equation}
\begin{array}{l}
E\left( a \right) = {E_{kin}} + {E_{int}} = \\
\int\limits_{ - \infty }^{ + \infty } {\left( {\begin{array}{*{20}{c}}
{\psi _A^*}&{\psi _B^*}
\end{array}} \right)\left( {\begin{array}{*{20}{c}}
0&{\left( {t - t'} \right) - itqa_0}\\
{\left( {t - t'} \right) + itqa_0}&0
\end{array}} \right)\left( {\begin{array}{*{20}{c}}
{{\psi _A}}\\
{{\psi _B}}
\end{array}} \right)} \,dx\\
 + \frac{1}{2}\alpha \int\limits_{ - \infty }^{ + \infty } {\left( {{{\left| {{\psi _A}} \right|}^4} + {{\left| {{\psi _B}} \right|}^4}} \right)dx}  = 2\left( {t - t'} \right) n + \frac{\alpha n^2}{{a\sqrt {6\pi } }}
\end{array}
\end{equation}

From this expression we see that the energy does not exhibit any extremum, neither maximum, nor minimum, as a function of the variational parameter. The wavepacket will therefore collapse (increasing its energy) until it becomes sufficiently small to invoke high wavevectors, so that our approximation becomes inexact and higher-order terms have to be taken into account.

Another trial function with its components $\psi_A\propto\exp(-x^2/(2a^2))$ and $\psi_B\propto x\exp(-x^2/(2a^2))$ gives a $1/a$ dependence for both kinetic energy and interaction energy, which does not allow a stable solution. Similar results can be obtained for displaced Gaussian trial functions and for the hyperbolic secant trial function. The reason for the instability is the linear dispersion of the Dirac equation at high wavevectors, which leads to $E\propto k \propto 1/a$ dependence of the kinetic energy, the same as that of the interaction energy.

\subsection{Full Hamiltonian}

We have calculated the variational energy for the full Hamiltonian using both hyperbolic secant (main text) and Gaussian trial wavefunctions, in order to obtain analytical equations for the variational parameters $a$ and $b$. The latter trial function is written as:

\begin{equation}
\psi_G \left( {x,a,b} \right) = \frac{{\sqrt n }}{{{{\left( {2a} \right)}^{1/2}}{\pi ^{1/4}}}}\left( {\begin{array}{*{20}{c}}
{\exp \left( { - \frac{{{{\left( {x - b} \right)}^2}}}{{2{a^2}}}} \right)}\\
{\exp \left( { - \frac{{{{\left( {x + b} \right)}^2}}}{{2{a^2}}}} \right)}
\end{array}} \right)
\end{equation}

We work with the Fourier transforms of the trial wavefunctions in order to calculate the kinetic energy. Keeping in mind that the two pseudospin components are respective complex conjugates, we can write:

\begin{widetext}
\begin{equation}
\begin{array}{c}
{E_{kin}} =  - \Re \int\limits_{ - \infty }^{ + \infty } {{{\left( {\psi _k^A\left( {k - \frac{\pi }{{{a_0}}}} \right) + \psi _k^A\left( {k + \frac{\pi }{{{a_0}}}} \right)} \right)}^*}} \left( {t' + t{e^{ - ik{a_0}}}} \right)\left( {\psi _k^B\left( {k - \frac{\pi }{{{a_0}}}} \right) + \psi _k^B\left( {k + \frac{\pi }{{{a_0}}}} \right)} \right)\,dk\\
 =  - \Re \int\limits_{ - \infty }^{ + \infty } {{{\left( {\psi _k^B\left( {k - \frac{\pi }{{{a_0}}}} \right) + \psi _k^B\left( {k + \frac{\pi }{{{a_0}}}} \right)} \right)}^2}} \left( {t' + t{e^{ - ik{a_0}}}} \right)\,dk
\end{array}
\end{equation}
\end{widetext}

The integration of the kinetic energy for the hyperbolic secant profile gives a cumbersome expression involving the hypergeometric function $_2 F_1$ that we do not show here. The Gaussian trial function gives a neat expression:
\begin{widetext}
\begin{equation}
{E_{Gkin}}\left( {a,b} \right) =  - 2{e^{ - \frac{{a_0^2 + 4{b^2} + \frac{{4{a^4}{\pi ^2}}}{{a_0^2}}}}{{4{a^2}}}}}n
\left( {\left( {1 + {e^{\frac{{{a^2}{\pi ^2}}}{{a_0^2}}}}} \right){e^{\frac{{a_0^2}}{{4{a^2}}}}}t' - \left( {{e^{\frac{{{a^2}{\pi ^2}}}{{a_0^2}}}} - 1} \right){e^{\frac{{{a_0}b}}{{{a^2}}}}}t} \right)
\end{equation}
\end{widetext}

Local maximum of the variational energy gives the following system of transcedental equations for $a$ and $b$:

\begin{widetext}
\begin{equation}
4\sqrt \pi  \left( {t{e^{\frac{{{a_0}b}}{{{a^2}}}}}\left( {4{a^4}{\pi ^2} + a_0^2{{\left( {{a_0} - 2b} \right)}^2}\left( {{e^{\frac{{{a^2}{\pi ^2}}}{{a_0^2}}}} - 1} \right)} \right) - 4t'{e^{\frac{{a_0^2}}{{4{a^2}}}}}\left( {a_0^2{b^2}\left( {1 + {e^{\frac{{{a^2}{\pi ^2}}}{{a_0^2}}}}} \right) - {a^4}{\pi ^2}} \right)} \right) - ana_0^2{e^{\frac{{a_0^2 + 4{b^2} + \frac{{4{a^4}{\pi ^2}}}{{a_0^2}}}}{{4{a^2}}}}} = 0
\end{equation}
\begin{equation}
\left( {{a_0} - 2b} \right){e^{\frac{{{a_0}b}}{{{a^2}}}}}\left( {{e^{\frac{{{a^2}{\pi ^2}}}{{a_0^2}}}} - 1} \right)t' + 2b{e^{\frac{{a_0^2}}{{4{a^2}}}}}\left( {1 + {e^{\frac{{{a^2}{\pi ^2}}}{{a_0^2}}}}} \right) t = 0
\end{equation}
\end{widetext}

Since the equation for $b$ does not depend on $n$, the displacement of the maxima of the two components is independent of the number of particles in the soliton. It is determined only by the difference $t'-t$ (characterizing the dimerization of the lattice), which is indeed confirmed by numerics.

\subsection{Existence of gap solitons}
The gap soliton solution exists only in a limited range of parameters, because its energy is limited by the size of the gap. The condition for the soliton existence is therefore different for GS (of the upper gap) and TGS (of the topological gap). Indeed, the upper (non-topological gap) is in reality limited by the energy difference between the quantized states of a single pillar $s$ and $p$: $E_p-E_s=3\pi^2\hbar^2/2m_{pol}L^2$, where $m_{pol}$ is the polariton mass. The central gap is limited by the difference in the tunneling coefficients $t-t'$, which is much smaller than $E_p-E_s$ (for the TB approximation to remain valid). The GS can therefore contain much more particles than the TGS:
\begin{equation}
\alpha n_{0,GS}\ll \frac{3\pi^2\hbar^2}{2m_{pol}L^2}
\end{equation}
and
\begin{equation}
\alpha n_{0,TGS}<\left|t-t'\right|
\end{equation}
where $n_0$ is the density at the soliton center.

\subsection{Soliton size and sublattice polarization degree}

\begin{figure}[tbp]
\includegraphics[scale=0.55]{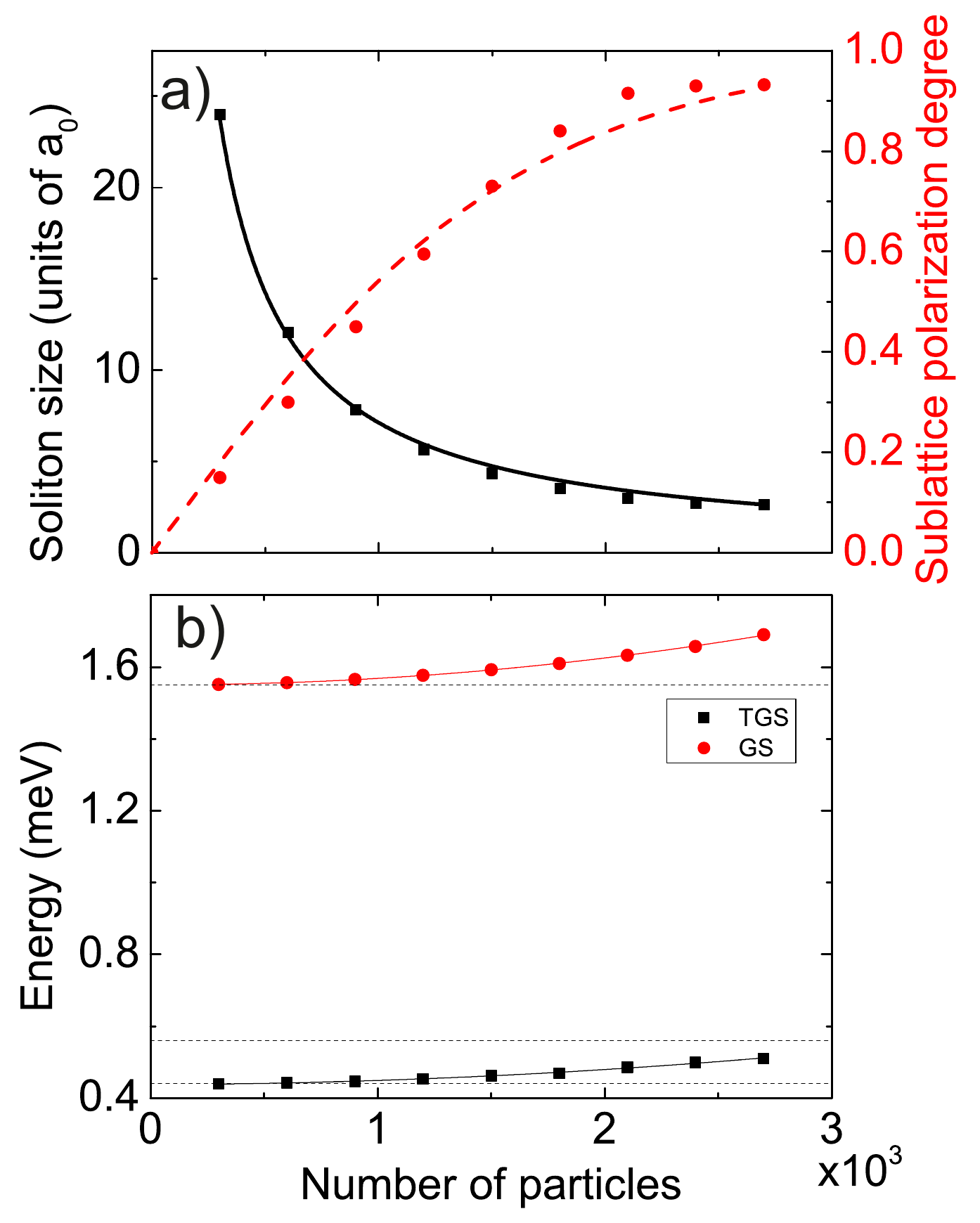}
\caption{ (Color online) a) Soliton width $a$ as a function of density $n$ obtained from numerical simulations (black dots) and found analytically (black line); Sublattice polarization  degree $\rho_{AB}$ as a function of density $n$ from numerics (red dots) and analytics (red line). b) Energy of TGS (black dots) and GS (red dots) with parabolic fits (black and red lines). The edges of the gaps are shown by dashed black lines. }
\label{figs1}
\end{figure}

We have checked that both gap solitons (TGS and GS) exhibit the typical dependence of their size $\xi=\hbar/\sqrt{\alpha n_0 m}$ and energy $E=\alpha n_0/2$ on the number of particles, where $n_0=|\psi (0)|^2$ is the density in the soliton center. The total number of particles in the soliton $n\propto n_0^2$. The width of the soliton as a function of the number of particles obtained from the full numerical simulation is shown in Fig. \ref{figs1}(a). Black dots show the results of the calculation and the black line is a fit with $1/n$. At the other hand, as predicted from our analytical calculations, the distance between the peaks in the two components $b$ does not exhibit any dependence on the number of particles, remaining approximately equal to 4 lattice periods for the lattice studied. Therefore, the sublattice polarization degree increases with $n$.

Figure \ref{figs1}(a) also shows the sublattice polarization degree $\rho_{AB\infty}$ at $x\to\infty$ as a function of the number of particles in the gap soliton extracted from the numerical solution of the full Gross-Pitaevskii equation (red points) together with a fit (red line) obtained by using the hyperbolic tangent function: $\rho_{AB}\propto \tanh{n}$. The numerical results confirm the expected analytical dependence.

Finally, Fig.  \ref{figs1}(b) shows the energy dependence on the total number of particles for both solitons, which grows quadratically, as expected. The energy of GS as a function of the total number of particles $n$ grows faster, because the soliton energy is directly proportional to the effective mass $E\propto m n^2$, and the effective mass at the non-topological gap is higher than the one of the topological gap.

\subsection{Numerical simulations}
In our solution of the Gross-Pitaevskii equation in the main text (Eq. 10, Fig. 4) we have used the parameters typical for exciton-polaritons: $m=5\times 10^{-5}m_0$, where $m_0=9.1\times 10^{-31}$ kg is the free electron mass, a periodic potential of 1 meV height with a period of 8 $\mu$m, and a 1D interaction constant $\alpha=20$ $\mu$eV/$\mu$m. These parameter can be optimized further in order to obtain faster oscillations.

In this section we also discuss the choice of the additional potential for the numerical simulations of the soliton behavior. The choice of the potential for the localized effective magnetic field is very important. Indeed, a localized potential $U_0$ on a single site (either A or B) induces an effective magnetic field in the $Z$ direction $\Omega=U_0/2$, but it also induces a potential acting on both pseudospin components $U_{eff}=U_0/2$. This is based on the following decomposition:

\begin{equation}
\left( {\begin{array}{*{20}{c}}
{{U_0}}&0\\
0&0
\end{array}} \right) = \frac{{{U_0}}}{2}\left( {\begin{array}{*{20}{c}}
1&0\\
0&1
\end{array}} \right) + \frac{{{U_0}}}{2}\left( {\begin{array}{*{20}{c}}
1&0\\
0&{ - 1}
\end{array}} \right) = \frac{{{U_0}}}{2}{{\bf{I}}_2} + \frac{{{U_0}}}{2}{\sigma _z}
\end{equation}

This potential, acting on both pseudospin components, will add to the energy of the soliton as a function of its position, which will therefore write:
\begin{eqnarray}
E&=&{E_{mag}}+U = \int \left( {\Omega  \cdot S} + \int U\left(x\right)\left|\psi\left(x\right)\right|^2\right)\,dx \nonumber \\
&\propto& \frac{\left({\tanh\left(b/a\right)\tanh \left( {{x_0}} \right)}+1\right)}{{{{\cosh }^2}\left( {{x_0}} \right)}}
\end{eqnarray}
This function does not allow the observation of two different types of behavior (oscillating and accelerating). Only one type of behavior will be observed, depending on the sign of the localized potential and independent of the initial position of the soliton (left or right).

On the other hand, a pure effective magnetic field acting on the pseudospin is composed of two potentials of opposite signs acting on A and B atoms. Although this field is a Delta function from the large-scale point of view, its internal structure (e.g. negative potential on the A atom, which is on the left of the B atom) explicitly breaks the spatial symmetry and induces a chiral behavior even for a trivial gap soliton, which itself is not chiral.

Therefore, to observe the difference between the two solitons, it is necessary to choose a proper potential, which would neither induce a chirality for the trivial gap soliton, nor suppress the chirality of the topological gap soliton. We have chosen a superposition of two Gaussian potentials of opposite signs and different spatial extension: 
\begin{equation}
U\left(x\right)=U_{+}e^{-x^2/w_{+}^2}-U_{-}e^{-x^2/w_{-}^2}
\end{equation}
where $U_{+}=5U_{-}$, $w_{+}=l_0/6$, $w_{-}=7l_0/6$, and $l_0$ is the size of a single site. In our calculations, the lattice period is $a_0=8$ $\mu$m and the site size is $l_0=3$ $\mu$m.
Experimentally, in polariton system, such potential can be achieved by varying the lateral size of the sites of the chain (for the negative contribution) and non-resonant optical pumping (for positive contribution).

Finally, we would like to note that this additional potential is introduced in the calculations at $t=20$ ps, in order to allow the soliton to stabilize its shape at the initial moments.

\begin{figure}[tbp]
\includegraphics[scale=0.7]{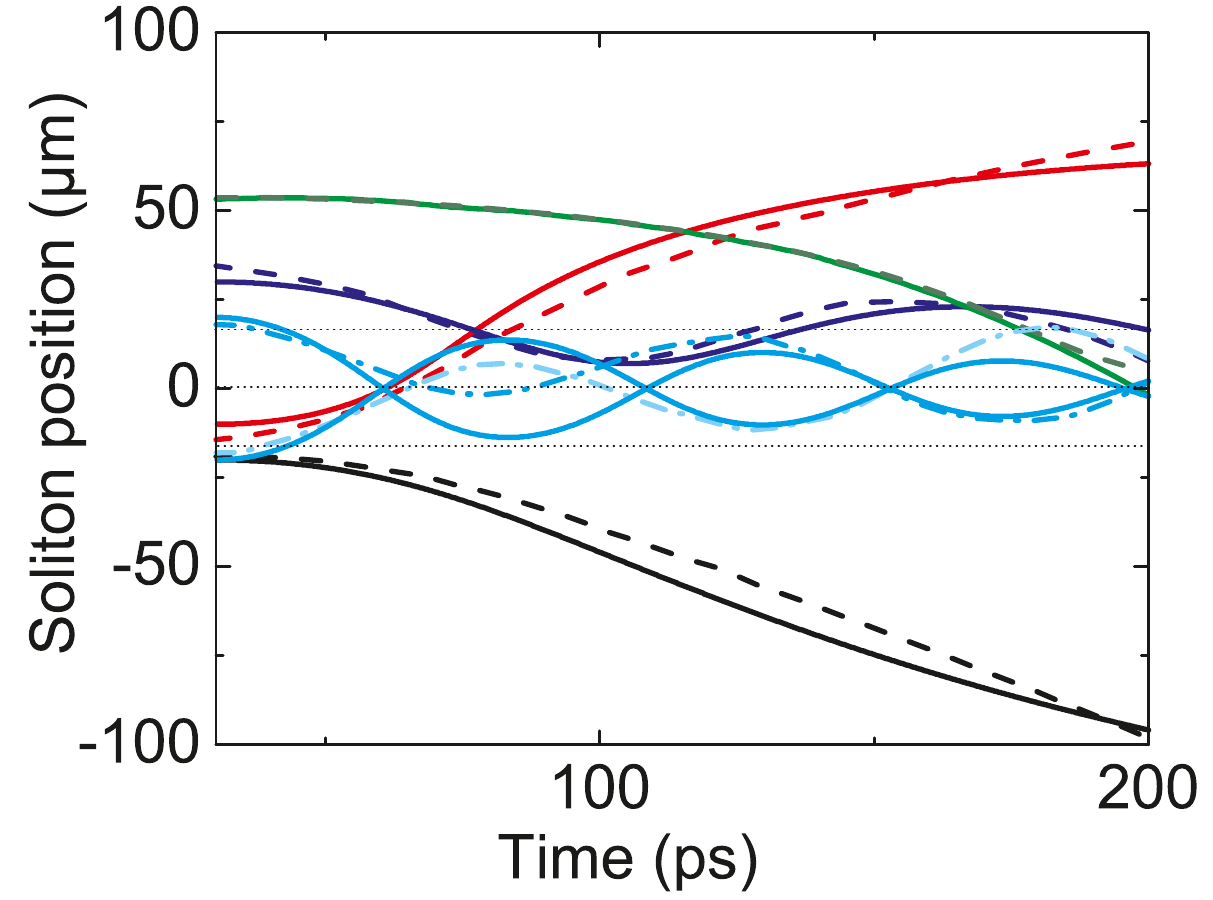}
\caption{ (Color online) Solid lines: soliton (center of mass) trajectories extracted from the full numerical simulation; dashed lines: solution of Hamilton's equations (main text). Black, red, green, navy: different families of TGS; light blue: GS. Dotted lines: extrema of the effective potential for TGS and GS.}
\label{figtraj}
\end{figure}

\subsection{Comparison of analytical and numerical trajectories}
Figure 2S shows the analytical trajectories (dashed lines), similar to those of Fig. 3 of the main text, but calculated using the Hamilton's equations with additional damping terms, which allows to obtain an optimal fit of the numerically calculated trajectories, also shown on the figure (solid lines). The latter were obtained by following the center of mass of the wavepacket. The dimensionless units used for analytical trajectories were rescaled to match the units of the numerical experiment.

The agreement between the full numerical simulation and Hamilton's equations confirms that solitons behave as well-defined particles. The chiral behavior of TGS (e.g. black and navy lines) becomes even more evident when compared with symmetric behavior of GS (light blue lines).

\bibliography{reference} 

\bibliography{reference} 

\end{document}